\documentclass[twocolumn,showpacs,preprintnumbers,amsmath,amssymb,prl]{revtex4-1}
\usepackage[latin1]{inputenc}
\usepackage{amsmath}
\usepackage{amsfonts}
\usepackage{amssymb}
\usepackage{graphicx}

\newcommand{\tref}[1]{Table~\ref{#1}}
\newcommand{\fermi}{G_\mathrm{F}}
\newcommand{\cm}{cm$^{-1}$}

\newcommand{\podd}{$\cal P$-odd~}
\newcommand{\ptodd}{$\cal P, T$-odd~}

\begin{document}

\author{T.~A.\ Isaev}
\email{timur.isaev.cacn@gmail.com}
\altaffiliation{present address: NRC KI - PNPI,
188300, Orlova Roscha 1, Gatchina, Russia}
\affiliation{Clemens-Sch{\"o}pf Institute,
             TU Darmstadt, Petersenstr. 22, 64287 Darmstadt, Germany}
\author{R.\ Berger}
\email{robert.berger@uni-marburg.de}
\altaffiliation{present address: Fachbereich Chemie, Philipps-Universit{\"a}t
Marburg, Hans-Meerwein-Str. 4, 35032 Marburg, Germany}
\affiliation{Clemens-Sch{\"o}pf Institute,
             TU Darmstadt, Petersenstr. 22, 64287 Darmstadt, Germany}

\title{Lasercooled radium monofluoride: A molecular all-in-one probe for new physics}

\begin{abstract}
The particular advantages of using the diatomic molecule radium monofluoride
(RaF) as a versatile molecular probe for physics beyond the Standard Model
are highlighted. i) RaF was previously suggested as being potentially
amenable to direct cooling with lasers. As shown in the present work,
RaF's energetically lowest electronically excited state is of ${}^{2}\Pi$
symmetry (in contrast to BaF), such that no low-lying ${}^{2}\Delta$ state
prevents efficient optical cooling cycles. ii) The effective electric field 
acting on the unpaired electron in the electronic ground state of RaF is 
estimated larger than in YbF, from which the best restrictions on the electron
electric dipole moment (eEDM) were obtained experimentally. iii) Favourable 
crossings of spin-rotational levels of opposite parity in external magnetic 
fields exist, which are important for the measurement of the nuclear anapole 
moment of nuclei with a valence neutron. Thus, RaF appears currently as one 
of the most attractive candidates for investigation of parity-odd as well
as simultaneously parity- and time-reversal-odd interactions in the realms 
of molecular physics.
\end{abstract}

\maketitle

\section*{Introduction}
The principle advantage of using heavy-atom, polar, diatomic molecules for
searching for space parity violation interactions (\podd interactions) and
simultaneously space parity and time reversal violating forces (\ptodd
forces) is known for more than 30 years. Nevertheless, only the latest
generation of molecular experiments finally surpassed their atomic
competitors in sensitivity to one of the most important \ptodd properties
of elementary particles, namely the permanent electric dipole moment of an
electron (eEDM) \cite{Hudson:11} (see also the recent report on preliminary
data for ThO which claims an improved restriction on the eEDM by almost an
order of magnitude \cite{acme:2013}).  Whereas molecules are ideally
tailored to create favourable, well-defined fields at the heavy nucleus,
the complexity of measurements with molecules is typically connected with
various systematic effects which can mimic \podd correlations.  This calls
for an active identification of promising molecular candidates that allow
different approaches to suppress systematic effects. In the search for an
eEDM, experimentally oriented research groups are currently focussing on
YbF \cite{Hudson:11}, PbO \cite{Demille:08}, ThO \cite{acme:11, acme:2013},
WC \cite{Lee:09}, PbF \cite{Shafer:06} molecules and HfH$^+$ molecular ion
\cite{Leanhardt:11}. In all these experiments, high-quality electronic
structure calculations are crucial both for the preparation stage of the
experiment and for subsequent interpretation of experimental data obtained
\cite{Titov:06amin}. 

Another set of pivotal molecular experiments is connected with attempts to
measure the nuclear anapole moment, a \podd electromagnetic form-factor
appearing in $I>0$ nuclei due to \podd nuclear forces. The only nucleus,
for which the anapole moment was successfully determined, is $^{133}$Cs.
In this experiment a vapour of Cs atoms was employed \cite{Wood:97}. The
results are apparently in disagreement with the earlier measurement on Tl
atom \cite{vetter:1995,ginges:2004}. Currently molecular experiments are
under development in Yale on BaF \cite{Demille:08} and in Groningen on SrF
\cite{vandenberg:2012}. 

Recently, we identified the open-shell diatomic molecule RaF as an
exceptionally suitable candidate for nuclear anapole moment measurements,
having on the one hand a high enhancement factor for nuclear spin-dependent
weak interaction and on the other hand offering potential for direct
cooling with lasers. In the present work we demonstrate that RaF presents
unique possibilities for measurements of \podd and \ptodd effects due to
favourable combinations of peculiarities of molecular electronic structure
and nuclear structure of radium isotopes.

\section*{Direct cooling of molecules with lasers}

We identified earlier a set of requirements on molecular electronic
structure for molecules being suitable for direct cooling with lasers
\cite{Isaev:10a}. Monofluorides of group II elements (e.g. BaF and RaF)
belong to the first class of molecules with highly diagonal Frank-Condon
matrix. One problem emerging, however, in lasercooling of BaF molecule
(isovalent to RaF) is connected with the existence of a metastable
$^2\Delta$ level, lying energetically below the $^2\Pi$ level involved in
the optical cooling loop.  Our previous electron correlation calculations
of the spectroscopic parameters in RaF indicated that the energetically
lowest electronically excited level is $^2\Pi$. We took now larger atomic
basis sets and active spaces of virtual molecular orbitals to investigate
the stability of the ordering of electronic levels. The results are
summarized in \tref{raf}. These show that even considerable alterations in the
parameters of the Fock-space relativistic coupled cluster (FS-RCC as
implemented in {\sc dirac} program package \cite{DIRAC:11})
calculations do not change the ordering of levels in RaF. On the other hand
FS-RCC, calculations of BaF with a basis set of similar quality as for
RaF and as large active spaces (see supplementary material) confirm the
first excited electronic level in BaF to be $^2\Delta$. Based on comparison
to experimental BaF results we estimate the accuracy of
$\tilde{T}_\mathrm{e}$ calculation for RaF to be within 1200 \cm (without
changing the ordering of the levels), $R_\mathrm{e}$ within about 0.1 $a_0$
and $\tilde{\omega}_\mathrm{e}$ about 60 \cm.

\begin{table}
\caption
 {Estimated molecular spectroscopic parameters for the
  electronic ground state $^2\Sigma_{1/2}$ and the first electronically
  excited states from FS-CCSD calculations of ${}^{226}$RaF 
  and ${}^{137}$BaF. 17 electrons are correlated in both cases: $2p_{1/2,3/2}$
  and $2s_{1/2}$ on fluorine, $5s_{1/2}$, $5p_{1/2,3/2}$ and  $6s_{1/2}$ 
  and Ba and $6s_{1/2}$, $6p_{1/2,3/2}$ and  $7s_{1/2}$ an Ra. 
  The results are given for two different 4-component all-electron uncontracted 
  basis sets on Ra and Ba atom:  Dyall's relativistic basis set and the RCC-ANO 
  basis set by Roos et al. The active space in FS-CCSD was restricted 
  by energy for Dyall's basis set up to 10 Hartree and for RCC-ANO up to 1000 
  Hartree (see supplementary material for computational details and references).
  To be directly comparable to experiment, the theoretical values given here
  for $\tilde{\omega}_\mathrm{e}$ have to be scaled by 
  $\sqrt{m_\mathrm{p}/\mathrm{u}}\approx
  1.0036$ as the mass of the proton $m_\mathrm{p}$ instead of the atomic mass 
  unit $\mathrm{u}$ was used to obtain $\tilde{\omega}_\mathrm{e}$ from the 
  fit of the computed potential curves to a Morse potential. We note that
  (co)variances of the fit parameters lead to relative 
  uncertainties of up to about 0.7~\% for the theoretical values of 
  $\tilde{\omega}_\mathrm{e}$ in RaF.
 } 
 \label{raf}
\begin{tabular}{lcccc}
& $R_\mathrm{e}/a_0$ & $\tilde{\omega}_\mathrm{e}/\mathrm{cm}^{-1}$  &
  $\tilde{T}_\mathrm{e}/(10^4 \mathrm{cm}^{-1})$   & $\tilde{D}_\mathrm{e}/(10^4 \mathrm{cm}^{-1})$ \\\hline
\multicolumn{5}{c}{RaF}\\
\multicolumn{5}{c}{Dyall basis set}\\
$^2\Sigma_{1/2}$                 & 4.24$^a$&  432$^a$&          & 3.21$^a$ \\
$^2\Pi_{1/2}$                    & 4.24$^a$&  428$^a$& 1.40$^a$& 3.13$^a$ \\
$^2\Pi_{3/2}+ ^2\Delta_{3/2}$     & 4.25    &  410    & 1.60    &           \\
$^2\Delta_{3/2}+ ^2\Pi_{3/2}$     & 4.25    &  432    & 1.64    &           \\
$^2\Delta_{5/2}$                 & 4.27    &  419    & 1.71    &           \\
$^2\Sigma_{1/2}$                 & 4.26    &  416    & 1.81    &           \\
\multicolumn{5}{c}{RCC-ANO basis set}\\
$^2\Sigma_{1/2}$                 & 4.29    &  431    &          & 4.26$^b$    \\
$^2\Pi_{1/2}$                    & 4.29    &  428    & 1.33    &           \\
$^2\Pi_{3/2}+ ^2\Delta_{3/2}$     & 4.31    &  415    & 1.50    &           \\
$^2\Delta_{3/2}+ ^2\Pi_{3/2}$     & 4.28    &  431    & 1.54    &           \\
$^2\Delta_{5/2}$                 & 4.30    &  423    & 1.58    &           \\
$^2\Sigma_{1/2}$                 & 4.32    &  419    & 1.67    &  \\\hline
\multicolumn{5}{c}{BaF}\\
\multicolumn{5}{c}{RCC-ANO basis set}\\
$^2\Sigma_{1/2}$                 & 4.15    &  456    &          & 4.67$^c$ \\
$^2\Delta_{3/2}$                 & 4.21    &  455    & 1.09    &    \\
$^2\Delta_{5/2}$                 & 4.21    &  455    & 1.13    &    \\
$^2\Pi_{1/2}$                    & 4.19    &  446    & 1.17    &    \\
$^2\Pi_{3/2}$                    & 4.19    &  444    & 1.24    &    \\
$^2\Sigma_{1/2}$                 & 4.23    &  455    & 1.42    &    \\
\multicolumn{5}{c}{Experimental data}\\
\multicolumn{5}{c}{(from Ref. \cite{effantin:1990} if not indicated otherwise)}\\
$^2\Sigma_{1/2}$                 & 4.09$^d$&  469$^d$&          &4.68 $\pm$ 0.07$^d$ \\
$^2\Delta$                       &         &  437    & 1.0940    &         \\
$^2\Pi$                          & 4.13$^e$&  437    & 1.1727    &         \\
$^2\Sigma$                       & 4.17$^e$&  424    & 1.3828    &         \\
\hline
\multicolumn{5}{l}{\footnotesize a) Ref.~\cite{Isaev:10a}, entries for $\tilde{\omega}_\mathrm{e}$
                                    of states $^2\Sigma_{1/2}$ and $^2\Pi_{1/2}$ }\\
\multicolumn{5}{l}{\footnotesize \phantom{a)} were erroneously exchanged in this reference;}\\
\multicolumn{5}{l}{\footnotesize b) Ref.~\cite{Isaev:13};
                                 c) Supplementary material of Ref.~\cite{Isaev:13}}\\
\multicolumn{5}{l}{\footnotesize                              
                                 d) Ref.~\cite{blue:1963};
                                 e) Ref.~\cite{nist-racl}}\\
\end{tabular}
\end{table}

\section*{Level crossing in magnetic field and sensitivity to the nuclear
anapole moment}

\podd effects in diatomic molecules can be greatly enhanced by shifting
levels of opposite parity to near-crossing with the help of external
magnetic fields \cite{flambaum:1985}. This idea is exploited in
\cite{Demille:08} for measurement attempts of the nuclear anapole moment in
BaF. One of the main problems in the suggested approach is to create highly
homogeneous magnetic fields in large volumes. Favorable values of magnetic
fields required to tune spin-rotational levels to near crossing would be
below 10~kG (1~T), as creation of larger magnetic flux densities require
special effort, for instance superconducting magnets. To estimate if fields
with $|B|<1$T suffice to create near level crossings in RaF we calculated
Zeeman splittings for spin-rotational levels (See Fig. 2 and Fig. 3 in
supplementary materials). Matrix elements of the spin-rotational
Hamiltonian in magnetic field were implemented as in
Ref.~\cite{kozlov:1991}. The following parameters of the spin-rotational
Hamiltonian were employed: rotational constant $B_\mathrm{e}=5689~$MHz (as
calculated from equilibrium structure), ratio of spin-doubling constant
$\Delta$ to $2B$, $\Delta/2B=0.97$ (as calculated within a four-component
Dirac--Hartree--Fock approach), components of the hyperfine tensor for
$^{225}$Ra nucleus $A_{\parallel}=-15100$ MHz and $A_\perp=-14800$ MHz
(\cite{Isaev:10a}, calculated with the two-component zeroth order regular approximation
(ZORA) approach as implemented in a modified version of the program package
{\sc turbomole} \cite{ahlrichs:1989}; values were already used for scaling
in Ref.~\cite{isaev:2012}, but not explicitly reported therein) and
components of $G$-tensor $G_{\parallel}=1.993$ and $G_\perp=1.961$ (crude
estimate based on results for HgH).  According to our calculation, the
first crossing of levels of opposite parity takes place at about 3~kG for
levels with the projection $F_z$ of the total angular momentum on the
direction of the magnetic field being $-3/2$.  A few more crossings take
place in fields up to 10~kG for levels with $F_z=-1/2$, thus providing
additional freedom for choosing the optimal experimental parameters.

To estimate the lowest possible flux of RaF molecules, which allows to
measure the anapole moment in RaF, one needs to find the ratio
$\frac{\Delta W}{W}$ with $W$ being the experimentally measured signal
proportional to the matrix element of the nuclear spin-dependent weak
interaction $W_\mathrm{a}$, and $\Delta W$ its experimental uncertainty.
The condition for meaningful measurement is $\frac{\Delta W}{W} < 1$. We
assume that the experimental scheme employed for measurement of the anapole
moment is analogous to the one suggested in \cite{Demille:08}.  According
to \cite{Demille:08} the maximal value of $\frac{\Delta W}{W}$ is
\begin{equation} \frac{\Delta W}{W} \simeq \frac{1}{2 \sqrt{2 N_0} t W},
\end{equation}
with $N_0$ being the total number of molecules ($N_0=F \tau$, where $F$ is
the {\it detected} molecular flux and $\tau$ is the total measurement time)
and $t$ the interaction time between molecule and external fields. Thus one
gets $F > 1/(8 W ^2 t^2\tau)$. The time for molecular trapping can reach a
few seconds \cite{Hoekstra:07}, so let $t=1$~s, $\tau=
1~\mathrm{h}=3600~\mathrm{s}$. To estimate $W$ we just scale the $W$ value
for Ba given in Ref.~\cite{Demille:08}, as $W^\mathrm{Ba}/W^\mathrm{Ra}
\simeq W_\mathrm{a}^\mathrm{Ba}/W_\mathrm{a}^\mathrm{Ra}$, and take
$W^\mathrm{Ra} \simeq 10W^\mathrm{Ba} = 50~\mathrm{Hz}$. The least required
flux of RaF is roughly then $F = 1/(8\cdot 2500 \cdot 1 \cdot
3600)~\mathrm{s}^{-1} = 1.4\cdot 10^{-8}~\mathrm{s}^{-1}$. In practice one
might expect trapping and detection of at least one molecule during an
experiment time of $\tau= 1~\mathrm{h}$ (see below), corresponding to
a flux of $2.8\cdot 10^{-4}~\mathrm{s}^{-1}$, which is a few orders of
magnitude higher than the least required flux of RaF. This
implies that it should be possible to perform successful measurements with
signals from {\it single} trapped RaF molecules.

\section*{Effective electric field acting on the unpaired electron in RaF}
One of the most important parameters in molecular experiments on eEDM is
the effective electric field $E_\mathrm{eff}$ acting on the unpaired
electron in the electronic ground state of the molecule of interest. This
field, however, can not be measured directly in experiment, but is
predicted from quantum chemical calculations.  We estimate here the
effective electric field acting on the unpaired electron in RaF by using
relations between matrix elements of different \podd and \ptodd operators
as it has been done in Ref.~\cite{kozlov:1985} by Kozlov and extended
recently in Ref.~\cite{dzuba:2011}. According to the semiempirical model of
Kozlov \cite{kozlov:1985} the relation between the parameter $W_\mathrm{s}$
of the \ptodd term and the parameter $W_\mathrm{a}$ of the \podd term in
the effective spin-rotational Hamiltonian is 
\begin{equation}
\label{eq:WstoWa}
W_\mathrm{s}/W_\mathrm{a} = Z 3 \gamma/(2 \gamma + 1),
\end{equation}
where $Z$ is the nuclear charge number and $\gamma = \sqrt{1-(\alpha
Z)^2}$.  As one can see from the \tref{all} this relation provides good
agreement with the results of explicit calculations of the $W_\mathrm{s}$
by a two-component ZORA generalized Hartree--Fock (GHF) method for BaF, YbF
and RaF molecules. On the other hand for HgH and CnH there exists a bigger
discrepancy between estimated and calculated $W_\mathrm{s}$.  This can be
attributed to the influence of the core-valence polarisation, which also
contributes considerably to the $W_\mathrm{a}$ values as noted in
\cite{isaev:2012}. To clarify the situation with the scalar \ptodd
interaction in HgH and CnH one needs high-precision correlation
calculations, similar to those in \cite{Isaev:04}.

Matrix elements $M_\mathrm{SPT}$ of the scalar \ptodd interaction and
matrix elements $M_\mathrm{EDM}$ of the coupling between eEDM and inner
molecular electric field are given in \cite{dzuba:2011}.  Relations between
these M.E.'s can be also expressed through relativistic enhancement factor
$R(Z)$ \cite{Moskalev:76}, which reflects the impact of relativistic
effects on molecular electronic structure (see e.g. \cite{isaev:2012} on
the influence of $R(Z)$ on $W_\mathrm{a}$ for a range of diatomic
molecules). The relation reads as
\begin{widetext}
\begin{eqnarray}
\frac{C_\mathrm{SP}}{(d_\mathrm{e}/(e a_0))} \frac{M_\mathrm{EDM}}{M_\mathrm{SPT}}=
\frac{(1-0.283\alpha^2Z^2)^2}{(1-0.56\alpha^2Z^2)}\frac{16\sqrt{2}\pi Z\alpha}{3 A 
(\fermi/(E_\mathrm{h}~a_0^3))}
\frac{2\gamma+1}{\gamma(1+\gamma)(4\gamma^2-1)} \frac{1}{R(Z)}.
\end{eqnarray}
\end{widetext}
In the equation above $G_\mathrm{F}$ is Fermi's constant, $A$ the atomic
mass number, $E_\mathrm{h}$ the Hartree energy and $a_0$ the Bohr radius.
$C_\mathrm{SP}$ and $d_\mathrm{e}$ are the effective constant of the scalar
\ptodd interaction and the electron electric dipole moment, respectively.
Herein the proper coefficient (namely 0.283) in front of $\alpha^2Z^2$ is
used in the numerator \cite{dzuba-priv:13}, instead of the misprinted one
(0.375) in \cite{dzuba:2011}. For the relation between $E_\mathrm{eff}$
and $W_\mathrm{s}$ one obtains then
\begin{equation}
\label{eq:WstoEeff}
E_\mathrm{eff}=-{\Omega}\frac{C_\mathrm{SP}}{d_\mathrm{e}}\frac{M_\mathrm{EDM}}{M_\mathrm{SPT}}
\frac{A}{Z}W_\mathrm{s},
\end{equation}
in which the quantum number $\Omega$ of the projection of the electron
total angular momentum on the internuclear axis is in the present work always 
taken equal to $1/2$.

It is interesting to note that in experiments with one kind of molecules
the scalar \ptodd interaction is indistinguishable from the eEDM effect --
they both contribute to the \ptodd electron paramagnetic resonance signal.
One can, however, disentangle the contributions by taking data from
experiments with different molecules (or molecule and atom) as
proposed in \cite{dzuba:2011} for eEDM measurements in Tl and YbF.

Taking into account the above relations one can easily estimate the
effective electric field acting on the electron in the ground $^2\Sigma$
state of RaF. The results are given in \tref{all}. The accuracy of such an
estimate is not high, but sufficient to identify RaF as promising candidate
for eEDM measurements.


\begin{table}
\caption
 {Calculated and estimated (in parentheses, absolute value only, from
  Eq.~\ref{eq:WstoWa}) $W_\mathrm{s}$ parameter of the $\cal P, T$-odd scalar 
  interaction  (in kHz) and the effective field on the electron 
  $|E_\mathrm{eff}|$ (in units of
 $10^{24}~\mathrm{Hz}~e^{-1}~\mathrm{cm}^{-1}$) for open-shell
  diatomic molecules. The value of $|E_\mathrm{eff}|$ is obtained via
  Eq.~\ref{eq:WstoEeff} from the numerically calculated $W_\mathrm{s}$ 
  parameter, the value in parentheses from the estimated $|W_\mathrm{s}|$
  value. For comparison, $|E_\mathrm{eff}|$ as calculated 
  {\it ab initio} previously is provided in square brackets.
}
\begin{tabular*}{\linewidth}{lrrrrr}
        & $Z$ &  & \multicolumn{1}{c}{$|W_\mathrm{a}|$/Hz$^{a)}$} 
                   & \multicolumn{1}{c}{$W_\mathrm{s}$/kHz} 
                     &
\multicolumn{1}{c}{$|E_\mathrm{eff}|$} \\
\hline
BaF     &  56 &     &  1.9$\times$10$^{2}$  &   $-$8.5 (12)              
&  1.3 (1.8) [1.9$^{b)}$]  \\ 
YbF     &  70 &     &  6.1$\times$10$^{2}$  &  $-$41 (38)              
&  4.4 (4.1) [6.0$^{c)}$]  \\ 
RaF     &  88 &     &  2.1$\times$10$^{3}$  & $-$15$\times$10$^{1}$ (13$\times$10$^{1}$)   
&  11 (9.5) \\ 
\hline
HgH     &  80 &     &  2.0$\times$10$^{3}$  & $-$38$\times$10$^{1}$ (19$\times$10$^{1}$)     
&  32 (~16) \\
CnH     & 112 &     &  3.1$\times$10$^{4}$  & $-$87$\times$10$^{2}$ (35$\times$10$^{2}$)    
&  746 (300) \\\hline
\multicolumn{6}{l}{\footnotesize a) Results of Ref.~\cite{isaev:2012}; 
b) Ref.~\cite{kozlov:1997}; c) Ref.~\cite{mosyagin:1998}}\\
\end{tabular*}
\label{all}
\end{table}
Another attractive feature of measurements with Ra nuclei is that there
exist also a nuclear mechanism enhancing \podd and \ptodd effects in certain
Ra isotopes. According to \cite{Auerbach:96} and \cite{Spevak:97} the Schiff
moment in nuclei possessing octapole deformation is enhanced by about
$10$ to $100$ times. The mechanism is similar to the one in
diatomic and chiral molecules: enhancement is reached due to
closeness of rotational levels of opposite parity. As a result
the estimated Schiff moments in $^{225}$Ra and $^{223}$Ra isotopes are
equal to 300 and 400 (in units $\eta~10^8~\mathrm{e}~\mathrm{fm}^3$,
where $\eta$ is effective nucleon-nucleon \ptodd force constant),
respectively, whereas for $^{199}$Hg for example it is only $-1.4$ (data 
taken from Ref.~\cite{Spevak:97}).

\section*{Production of RaF}
\label{prod} 
Besides the possible routes discussed in Ref.~\cite{Isaev:10a}, we propose in
Ref.~\cite{Isaev:13} to produce neutral RaF via RaF$^+$, which is subsequently 
neutralised by charge exchange in collision with a suitably chosen collision 
gas or by interaction with surfaces that provide the adequate work function 
for an iso-enthalpic electron transfer.  RaF$^+$ can in turn be formed 
in reactive collisions of radium ions with a suitable fluorine containing 
compound.

\section*{Conclusion}
We demonstrated various special properties of RaF that render it a
versatile molecular laboratory for studying a wide range of physical
phenomena, from laser cooling to physics beyond the Standard Model. A
unique combination of rovibronic and nuclear structure features renders RaF
particularly attractive for further experimental study.  For the first time
the parameter of the scalar \ptodd interaction $W_\mathrm{s}$ is calculated
with the accounting for spin-polarisation for the molecules RaF, HgH and
CnH. The authors are grateful to D. DeMille, V. Flambaum, M. Kozlov and S.
Hoekstra for discussion.

\bibliographystyle{apsrev4-1}
%
\end{document}